\documentclass[aps, prl, 10pt, twocolumn, superscriptaddress, noshowpacs, preprintnumbers, longbibliography, bibnotes,hyperref,floatfix]{revtex4-1}

\usepackage[colorlinks=true,breaklinks=true]{hyperref}
\hypersetup{allcolors=[rgb]{0.0 0.0 1.0},linkcolor=[rgb]{0.75 0.05 0.05}}
\usepackage[dvipsnames]{xcolor}
\usepackage{mathtools}
\usepackage{amsmath}
\usepackage{amsfonts}
\usepackage{amssymb}
\usepackage{bbold}
\usepackage{multirow}
\usepackage{bm}
\usepackage{hyperref}
\long\def\exclude#1{}
\usepackage{orcidlink}

\newcommand{\bp}{{\bf p}}

\newcommand{\RNS}{R_{\rm NS}}
\newcommand{\Rprog}{R_{\rm prog}}

\begin{document}

\title{Energy transfer by feebly interacting particles in supernovae: the trapping regime}

\author{Damiano F.\ G.\ Fiorillo \orcidlink{0000-0003-4927-9850}}
\email{damianofg@gmail.com}
\affiliation{Deutsches Elektronen-Synchrotron DESY,
Platanenallee 6, 15738 Zeuthen, Germany}

\author{Tetyana Pitik \orcidlink{0000-0002-9109-2451}}
\email{tetyana.pitik@berkeley.edu}
\affiliation{Department of Physics, University of California, Berkeley, Berkeley, CA 94720, USA}
\affiliation{Institute for Gravitation and the Cosmos, The Pennsylvania State University, University Park PA 16802, USA}

\author{Edoardo Vitagliano \orcidlink{0000-0001-7847-1281}} 
\email{edoardo.vitagliano@unipd.it}
\affiliation{Dipartimento di Fisica e Astronomia, Universit\`a degli Studi di Padova, Via Marzolo 8, 35131 Padova, Italy}
\affiliation{Istituto Nazionale di Fisica Nucleare (INFN), Sezione di Padova, Via Marzolo 8, 35131 Padova, Italy}

\begin{abstract}

Feebly interacting particles, such as sterile neutrinos, dark photons, and axions, can be abundantly produced in the proto-neutron star (PNS) formed in core-collapse supernovae (CCSNe). These particles can decay into photons or charged leptons, depositing energy outside the PNS. Strong bounds on new particles can thus be derived from the observed luminosity of CCSNe, with even tighter bounds obtained from low-energy SNe observations.
For the first time we highlight that, at sufficiently large couplings, particle production \textit{outside} the PNS must also be considered. Using the prototypical case of axions coupling to two photons, we show that at large couplings the energy transfer from PNS to its surroundings is diffusive rather than ballistic, substantially reducing the deposited energy. Our findings have implications for the parameter space of particles probed in beam dump experiments and for dark matter models involving a sub-GeV mediator.

\end{abstract}

\date{\today}

\maketitle

{\bf\textit{Introduction.}}---Finding feebly interacting particles (FIPs) is the goal of many laboratory, astrophysical, and cosmological searches~\cite{Essig:2013lka,Agrawal:2021dbo,Abdullahi:2022jlv,Berryman:2022hds,Baryakhtar:2022hbu}. After they are produced---be it by a beam interacting with a target, in the hot and dense core of proto-neutron stars (PNSs), or in the early universe---sterile neutrinos, dark photons, and axions with masses below the GeV scale can propagate and decay into Standard Model particles. Decay lengths comparable to the meter scale are constrained by beam-dump experiments (see e.g.~\cite{CHARM:1985anb,Riordan:1987aw,Blumlein:1990ay,NA64:2020qwq,Dolan:2017osp,Bauer:2017ris,Bauer:2018onh,Magill:2018jla,Krnjaic:2019rsv,Bolton:2019pcu,Calibbi:2020jvd,Capozzi:2023ffu,Knapen:2024fvh,Li:2025yzb}), whereas novel particles that survive on cosmological scales would have an effect on big bang nucleosynthesis or the cosmic microwave background (see e.g.~\cite{Redondo:2008ec,Cadamuro:2011fd,Fradette:2014sza,Berger:2016vxi,Depta:2019lbe,Fradette:2018hhl,Escudero:2019gzq,Depta:2020wmr,Depta:2020zbh,Mastrototaro:2021wzl,Sabti:2020yrt,Boyarsky:2020dzc,Langhoff:2022bij,DEramo:2024lsk,Akita:2024nam}). Astrophysical transients cover the gap between these observables: the remnants of neutron star mergers~\cite{Diamond:2021ekg,Diamond:2023cto,Dev:2023hax}, core-collapse supernovae (CCSNe)~\cite{Oberauer:1993yr,Davidson:2000hf,Giannotti:2010ty,Kazanas:2014mca,Chang:2016ntp,Jaeckel:2017tud,Chang:2018rso,Brdar:2020quo,Croon:2020lrf,Camalich:2020wac,Lucente:2020whw,Caputo:2021rux,Calore:2021lih,Caputo:2022mah,Caputo:2022rca,Ferreira:2022xlw,Hoof:2022xbe,Lella:2022uwi,Fiorillo:2022cdq,Akita:2022etk,Caputo:2022rca,Fiorillo:2023cas,Fiorillo:2023ytr,Diamond:2023scc,Lella:2023bfb,Carenza:2023old,Akita:2023iwq,Chauhan:2023sci,Lella:2024dmx,Fiorillo:2024upk,Telalovic:2024cot,Benabou:2024jlj,Chauhan:2024nfa,Alda:2024cxn}, or yet rarer events such as hypernovae~\cite{Caputo:2021kcv} can abundantly produce FIPs with masses as heavy as the GeV scale. 

\begin{figure}[h!]
    \centering
    \includegraphics[width=0.5\textwidth]{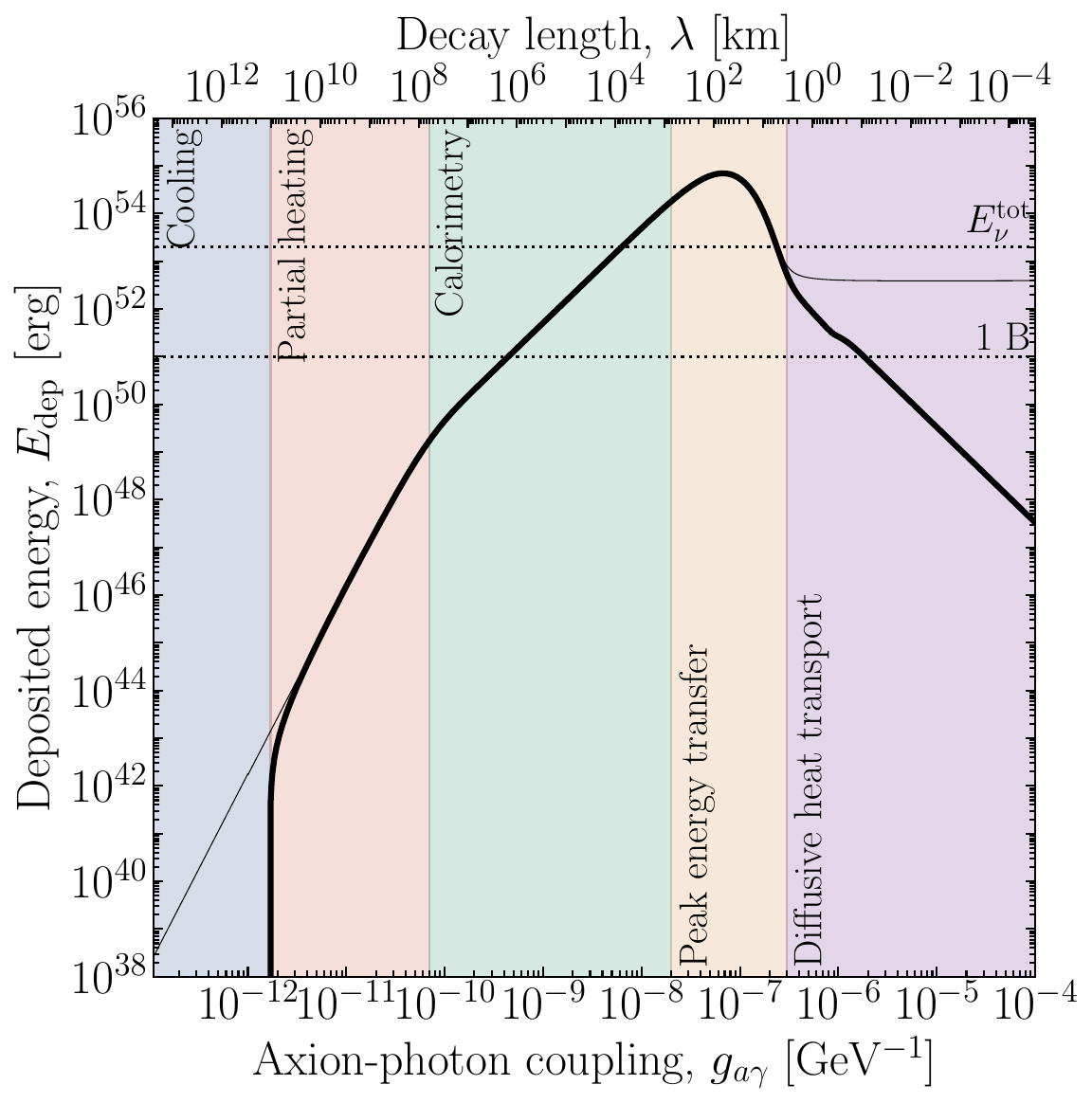}
    \caption{Energy deposited outside the PNS across different regimes of interaction strength. We choose an axion mass $m_a=80\,\rm MeV$ and a progenitor radius $R_1=3\times 10^{12}\,\rm cm$. The thin line accounts only for axions produced within the PNS (Eq.~\ref{eq:wrong_eq_1}), the thick line includes production outside of the PNS (Eq.~\ref{eq:e_dep_correct}). CCSNe can constrain $E_{\rm dep}\gtrsim 1\,\rm B$; above $E^{\rm tot}_\nu$, the total energy emitted in neutrinos~\cite{Fiorillo:2022cdq}, post-processing the SN model is certainly unphysical. The top axis shows the decay length for an axion with energy $E_a=300\,\rm MeV$.}
    \label{fig:fig1}
\end{figure}

A CCSN consists of a central PNS enveloped by an extended mantle. Axions decaying into photons would impact the observed features of CCSNe in different ways.  
If the mean free path of axions produced in the PNS is much larger than the progenitor radius, $\lambda\gtrsim R_{\rm prog}$, bounds are obtained from the putative shortening of the neutrino signal detected from SN~1987A~\cite{Raffelt:1996wa,Lucente:2020whw,Caputo:2021rux}, from x-ray~\cite{Diamond:2023scc} and $\gamma$-ray observations of SN~1987A~\cite{Jaeckel:2017tud,Caputo:2021rux,Hoof:2022xbe}, as well as from the diffuse gamma-ray flux observed at Earth, since the decay of axions produced by all past SNe would contribute to the latter~\cite{Calore:2020tjw,Caputo:2021rux}. For larger decay lengths, corresponding to smaller masses, massive stars that have not yet exploded are optimal sources, since axions would be produced with smaller Lorentz boost~\cite{Candon:2024eah}.
If, on the other hand, the decay length is shorter than the radius of the progenitor, $\lambda\lesssim R_{\rm prog}$, the axion can decay in the material surrounding the core, transporting energy and resulting in an exceedingly energetic explosions. The requirement that the energy deposited in the mantle of normal CCSNe and low-energy SNe (LESNe) is smaller than the observed values, respectively $1\,\rm B \,(bethe)=10^{51}\, erg$ and $0.1\,\rm B$, leads to strong constraints on the existence of novel radiatively decaying particles~\cite{Falk:1978kf,Sung:2019xie,Caputo:2022mah}. Ultimately, this is a transport of energy from the hot PNS to its surroundings, mediated by axions. While the focus is often on the free-streaming regime, i.e. on axions that interact so feebly that they escape far from the PNS after being produced, here we answer the question: how \textit{small} does the decay length need to be to avoid lighting up normal CCSNe and LESNe?

\begin{figure}
    \centering
    \includegraphics[width=0.5\textwidth]{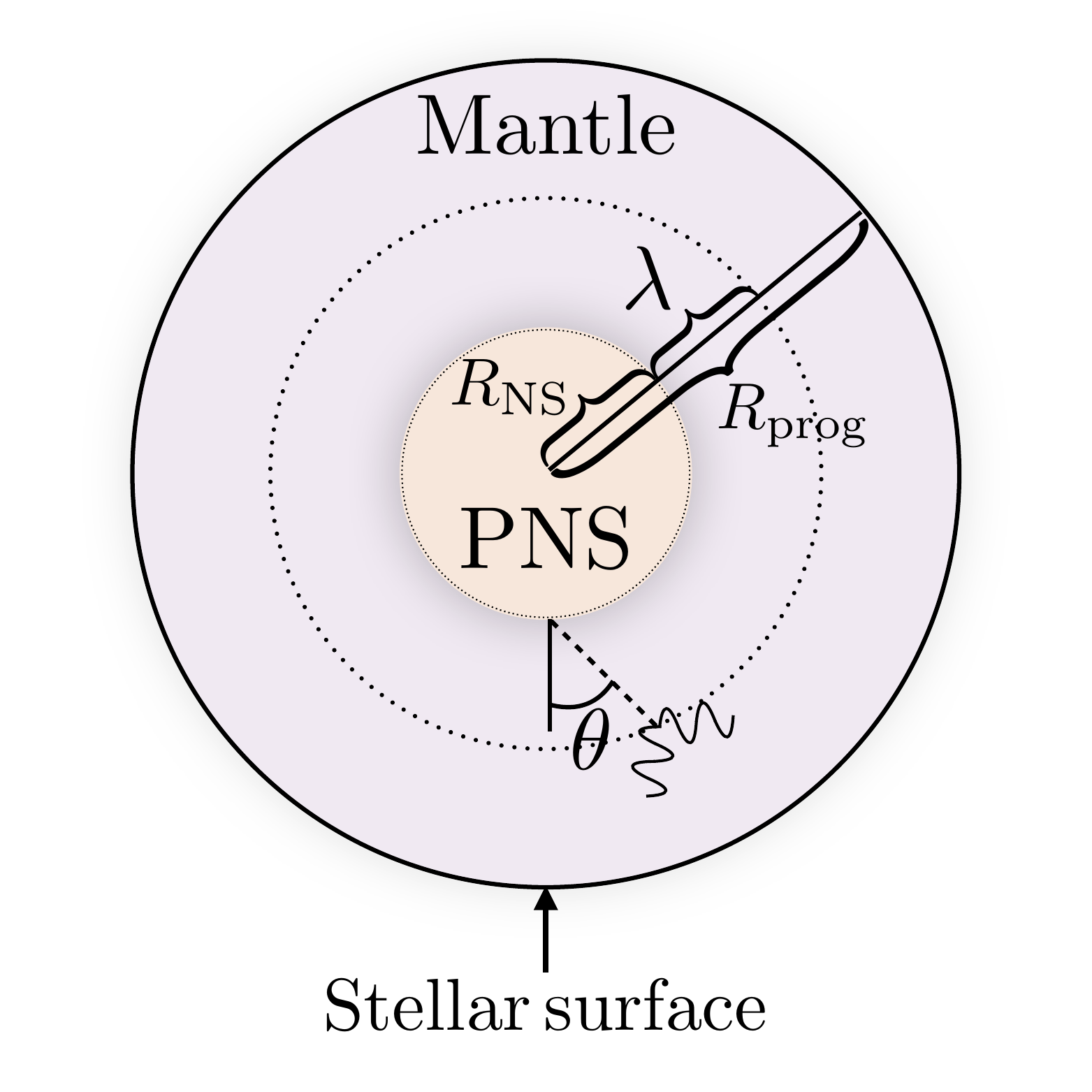}
    \caption{ Schematic geometry of the diffusive heat transfer in the trapping regime: axions are nearly in thermal equilibrium outside the PNS, with a small diffusive flux driven by the temperature gradient; at the axiosphere, where $R\sim \lambda$, they decay, dissipating all the energy locally into heat.}
    \label{fig:scheme}
\end{figure}

Fig.~\ref{fig:fig1} summarizes the different transport regimes. At very low axion-photon coupling, the progenitor undergoes \textit{cooling} rather than heating, as the energy lost through direct axion emission exceeds the energy transported from the PNS into the medium. At higher couplings, a fraction of the axions from the PNS decay within the progenitor, surpassing the cooling rate and leading to \textit{partial heating} of the star. When the axion mean free path becomes comparable to the progenitor radius, the system enters a \textit{calorimetric} regime, where all the energy produced in the PNS is deposited in the progenitor. As the mean free path decreases further and approaches the PNS radius, energy transfer reaches its \textit{peak efficiency}, with axions acting as the most effective transport mechanism. Finally, once axions thermalize within the PNS, transport from the PNS to the star becomes \textit{purely diffusive}. At this stage, axions remain in thermal equilibrium outside of the PNS, with a net heat flux determined by their thermal conductivity; this heat flux will ultimately dissipate into photons at larger radii, when the temperature becomes so small that the axions decay completely. In this regime, the heat transport is conceptually similar to the neutrino-delayed explosion mechanism, except that axions dissipate their energy by decay rather than scattering, and so the entire heat flux is ultimately deposited at radii much smaller than the gain radius. The geometry of this newly identified regime is schematized in Fig.~\ref{fig:scheme}. Our key insight is that this regime, rather than the kinetic one, is the relevant trapping mechanism for new particles with masses below a few hundred MeV.

{\bf\textit{Energy deposition at large couplings.}}---
In previous works~\cite{Sung:2019xie,Caputo:2022mah}, energy deposition has been determined as the difference between the total energy of axions emitted from the PNS (assumed to have a radius of $R_{\rm NS}=20\,\rm km$), and the total energy these axions retain beyond the progenitor radius. These two contributions can be computed as follows. At a given radius $r$, we assume the energy emission rate is $dL/dr$, while the absorption rate per unit length, due to either decay or scattering, is $1/\lambda(r)$, where $\lambda(r)=v_a/\tilde{\gamma}_a(r)$ is the mean free path expressed in terms of the reduced absorption coefficient (we review its definition in the Supplemental Material (SupM)~\cite{supplementalmaterial}), and provide its explicit expression in the case of axion-photon coupling).  The total energy emitted per unit time from the PNS is then
\begin{equation}
    \frac{dL^{(1)}_{\rm PNS}}{dE_a}=\int_0^{\RNS} dR \frac{dL}{dE_a dR} \int \frac{dx}{2} e^{-\tau(R,\RNS)}.
\end{equation}
Here, $E_a$ is the axion energy, $x=\cos\theta$ represents the angle between the axion trajectory and the radial direction, and $\tau(R,\RNS)$ is the optical thickness for an axion traversing the distance from $R$ to $\RNS$. Although it is more natural to express this in terms of the distance $s$ along the axion trajectory from its production point, the absorption rate $\lambda(r)$ depends on the radius $r = \sqrt{R^2 + s^2 + 2Rsx}$, making it more convenient to express it as a function of $r$. Thus, we obtain
\begin{equation}
    \tau(R,\RNS)=\begin{cases}
        
    \int_R^{\RNS}\frac{r dr}{\lambda(r) \sqrt{r^2-R^2\sin^2\theta}},\;x>0 \\
    \left[\int_{R\sin\theta}^R+\int_{R\sin\theta}^{\RNS}\right]\frac{r dr}{\lambda(r) \sqrt{r^2-R^2\sin^2\theta}},\;x<0 .
    \end{cases}
\end{equation}
Since we aim to capture the transition from free streaming to trapping, when axions are near-isotropic close to the PNS surface, we have to retain the angular dependence and cannot assume pure radial propagation. The differential luminosity of axions emitted from the PNS that successfully escape the progenitor is given by
\begin{equation}
    \frac{dL^{(2)}_{\rm PNS}}{dE_a}=\int_0^{\RNS} dR \frac{dL}{dE_a dR} \int \frac{dx}{2} e^{-\tau(R,\Rprog)},
\end{equation}
where $\Rprog$ is the progenitor radius. According to our qualitative arguments above, we would expect that the rate of energy extracted from the PNS and deposited in the progenitor is
\begin{equation}\label{eq:wrong_eq_1}
    E^{\rm PNS}_{\rm dep}=\int dt \left[L^{(1)}_{\rm PNS}-L^{(2)}_{\rm PNS}\right],
\end{equation}
which conceptually coincides, e.g., with Eq.~(1) of Ref.~\cite{Caputo:2022mah}.

So far, we have assumed the axions to be produced exclusively in the PNS. However, this is generally incorrect and becomes inconsistent for short mean free paths. If the particle decays rapidly enough to thermalize outside of the PNS, equally efficient production processes must also exist in that region. In the limit of an arbitrarily small mean free path, $L^{(2)}_{\rm PNS}$ vanishes, while $L^{(1)}_{\rm PNS}$ saturates to a constant value---the Stefan-Boltzmann luminosity from the PNS surface. However, our previous description of energy deposition as a transport process makes it clear that this cannot be correct, as a species with a mean free path shorter than that of neutrinos cannot transport energy more efficiently than them. The key insight is that Eq.~\eqref{eq:wrong_eq_1} must consistently include the negative contribution from axions produced \textit{outside} the PNS; it is convenient to consider separately the axions produced here and decaying either outside of the progenitor or within the PNS. The first contribution is easiest to express, as it is identical to $L^{(2)}_{\rm PNS}$ but integrated only outside the PNS
\begin{equation}
    \frac{dL^{(2)}_{\rm prog}}{dE_a}=\int_{\RNS}^{\Rprog} dR \frac{dL}{dE_a dR} \int \frac{dx}{2} e^{-\tau(R,\Rprog)}.
\end{equation}
Let us now consider the axions produced outside of the PNS and decaying inside the PNS. If an axion is produced at a distance $R>\RNS$, it can only decay inside the PNS if it is emitted at an angle $x<-\sqrt{1-\RNS^2/R^2}$; here we neglect any distortion of the axion trajectory due to gravity. The energy extracted from the progenitor due to these axions is therefore
\begin{align}
    \frac{dL^{(1)}_{\rm prog}}{dE_a}=\int_{\RNS}^{\Rprog} &dR\frac{dL}{dE_a dR}
    \nonumber
    \\
    &\times\int_{-1}^{-\sqrt{1-\frac{\RNS^2}{R^2}}} \frac{dx}{2}e^{-\tau_1(R)}(1-e^{-\tau_2}), 
\end{align}
where
\begin{equation}
    \tau_1(R)=\int_{\RNS}^R \frac{rdr}{\lambda(r) \sqrt{r^2-R^2\sin^2\theta}}
\end{equation}
is the optical depth accumulated from the production point to the point where the axion enters the PNS, and
\begin{equation}
    \tau_2=2\int_{R\sin\theta}^{\RNS}\frac{rdr}{\lambda(r) \sqrt{r^2-R^2\sin^2\theta}}
\end{equation}
is the optical depth accumulated along the part of the trajectory within the PNS. We have assumed that the SN profile is stationary, since the typical timescale over which the medium changes, a fraction of a second, is longer than the light-crossing time through the profile itself.

Therefore, our final expression for the energy deposited within the progenitor is
\begin{equation}\label{eq:e_dep_correct}
    E_{\rm dep}=\int dt \left[L^{(1)}_{\rm PNS}-L^{(2)}_{\rm PNS}-L^{(1)}_{\rm prog}-L^{(2)}_{\rm prog}\right].
\end{equation}

\begin{figure*}
    \includegraphics[width=0.95\textwidth]{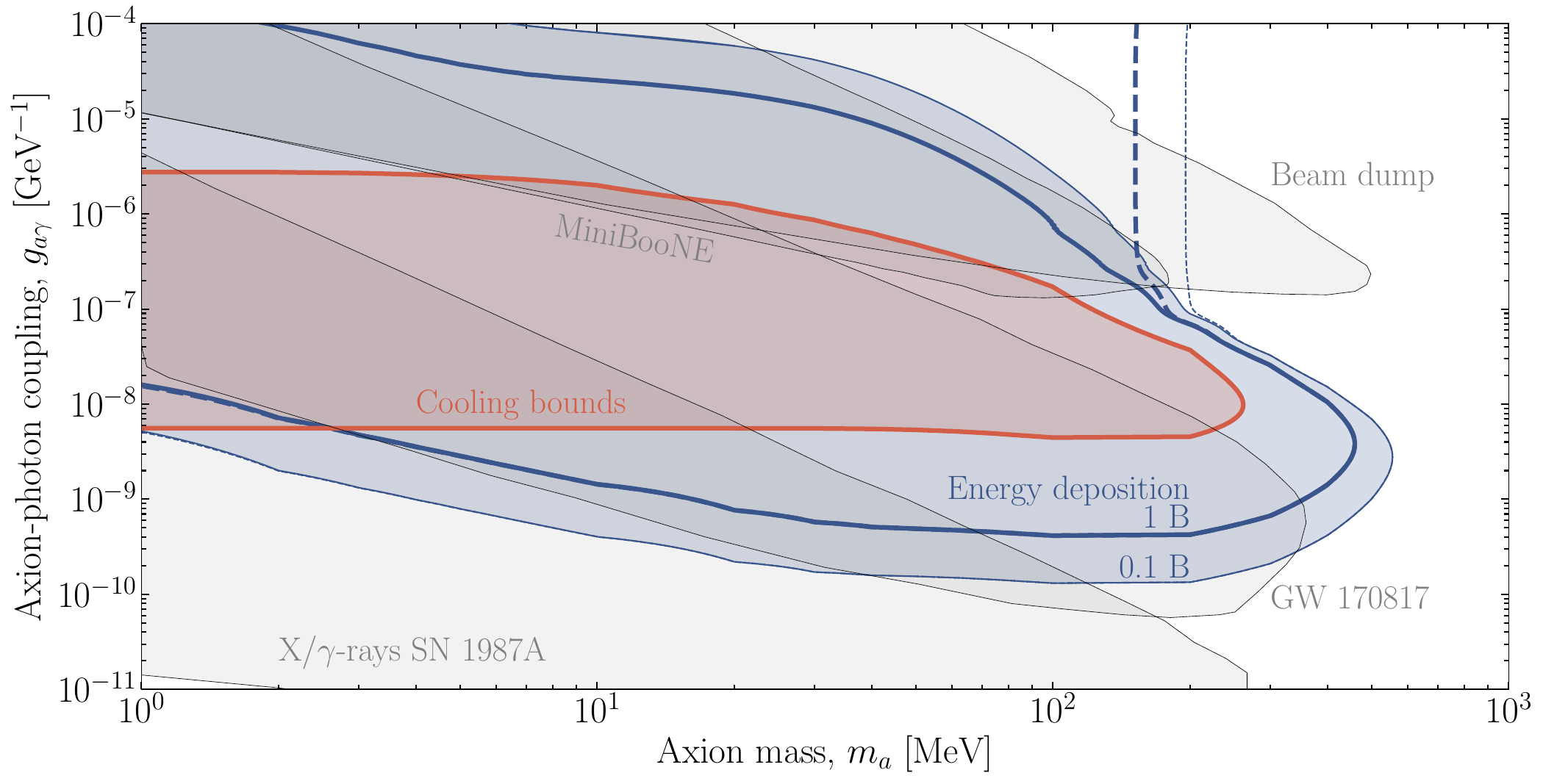}
    \caption{Energy-deposition and cooling constraints on axion-photon coupling. Similarly to Ref.~\cite{Caputo:2022mah}, energy-deposition constraints are shown both for a large progenitor ($\Rprog=5\times 10^{13}\,\rm cm$) with an explosion energy $E_{\rm dep}=0.1\,\rm B=10^{50} \, erg$, and for a smaller progenitor ($\Rprog=3\times 10^{12}\, \rm cm$) with an explosion energy $E_{\rm dep}=1\,\rm B$. For the latter, we also show with a dashed line the constraints that would be drawn using Eq.~\eqref{eq:wrong_eq_1}, neglecting particle production outside of the PNS. Previous Earth-based~\cite{CHARM:1985anb,Riordan:1987aw,Blumlein:1990ay,NA64:2020qwq,Dolan:2017osp,Capozzi:2023ffu} and astrophysical~\cite{Jaeckel:2017tud,Hoof:2022xbe,Diamond:2023cto,Diamond:2023scc} constraints are shown in gray. }\label{fig:fig2}
\end{figure*}

Eq.~\eqref{eq:e_dep_correct} encompasses all the regimes identified in Fig.~\ref{fig:fig1}; we use it for the axion coupling to photons, which is produced primarily by Primakoff conversion $\gamma N\to a N$ and photon-photon coalescence $\gamma\gamma\to a$. At small couplings, $L^{(1)}_{\rm PNS}$ and $L^{(2)}_{\rm PNS}$ are very similar, since the vast majority of axions decays outside the progenitor; their difference vanishes in proportion to $g_{a\gamma}^4$, since the fraction of axions decaying within the progenitor decreases as $\Rprog/\lambda\propto g_{a\gamma}^2$. Hence, Eq.~\eqref{eq:wrong_eq_1} would predict $E_{\rm dep}^{\rm PNS}\propto g_{a\gamma}^4$. However, in Eq.~\eqref{eq:e_dep_correct} the term $L^{(2)}_{\rm prog}$ grows as $g_{a\gamma}^2$ and must dominate at sufficiently small couplings; this is the cooling regime of Fig.~\ref{fig:fig1}. This regime is relatively unimportant for axion constraints. At higher couplings, $E^{\rm PNS}_{\rm dep}$ dominates, and the axions produced inside the PNS deposit part or all of their energy in its surroundings. Across the phases of partial heating, calorimetry, and peak energy transfer, Eq.~\eqref{eq:wrong_eq_1} correctly represents the deposited energy, with the transport being dominated by the energy emitted inside the PNS and deposited outside of it. However, when the axion reaches full thermalization at the PNS surface and the ``axiosphere''---the sphere at which axions decouple from the medium, with $\lambda(R)\lesssim  R$---moves outside of it,  Eq.~\eqref{eq:wrong_eq_1} would predict a constant thermal energy flux, corresponding to $L^{(1)}_{\rm PNS}$. However, this flux is canceled by $L^{(1)}_{\rm prog}$, so that the only residual flux is diffusive energy transport; axions have become part of the fluid that constitutes the medium, and can only contribute to its heating and cooling via their thermal conductivity. In fact, it is more convenient to determine this residual flow directly, rather than relying on Eq.~\eqref{eq:e_dep_correct}, which involves a near-cancellation that is difficult to capture numerically. To do so, we use the standard transport solution for the axion phase-space distribution $f$ at the PNS surface, which  locally depends only on $r$ and on the angle with the radial direction since $\lambda(\RNS)\ll \RNS$,
\begin{equation}
    f\simeq f^{\rm eq}-\lambda x \frac{\partial f^{\rm eq}}{\partial r},
\end{equation}
where $f^{\rm eq}=(e^{E_a/T}-1)^{-1}$ is the equilibrium distribution.
From here, we can compute the energy flux passing through the PNS surface as $q=\int f E_a  v_a x d^3\bp/(2\pi)^3$, with $\bp$ the axion momentum, so that the total rate of energy deposited outside the PNS is
\begin{align}\label{eq:diffusive_energy_transfer}
    \frac{dE_{\rm dep}}{dt}&=4\pi R_{\rm NS}^2 q
    \nonumber \\
    &=\frac{-2 R_{\rm NS}^2 \partial T/\partial z}{3\pi T^2}\int dE_a \frac{\lambda(E_a)E_a^2 p_a^2 e^{E_a/T}}{(e^{E_a/T}-1)^2},
\end{align}
where all quantities are evaluated at the PNS radius. The deposited energy drops as $g_{a\gamma}^{-2}$, and it smoothly connects with the numerical prediction of Eq.~\eqref{eq:e_dep_correct}, as visible in Fig.~\ref{fig:fig1}. Crucially, the heat flux we determine is extracted from the PNS, but in its immediate surroundings this energy remains in the form of axions, which are still in thermal equilibrium. This energy will ultimately be deposited when the medium temperature drops significantly below the mass of the axion, so that the entire axion population decays into photons and heats the medium.

Although our focus is on the energy deposition argument, a similar issue applies to the well-established cooling bounds, based on the argument that a rapid PNS cooling would shorten the duration of the neutrino burst~\cite{Mayle:1987as,Turner:1987by,Burrows:1988ah,Mayle:1989yx,Burrows:1990pk,Raffelt:1996wa} (see, however, Ref.~\cite{Fiorillo:2023frv} for a critical discussion). In practice, one often requires that the axion luminosity \textit{at infinity} exceeds the neutrino luminosity. However, only the energy extracted directly from the PNS can influence the neutrino signal, making the axion luminosity at infinity an unreliable proxy for this effect. For decaying particles, this is particularly clear, as the axion luminosity at infinity vanishes -- all particles eventually decay. 
Thus, here we use as a proxy the energy lost by the PNS $L=L^{(1)}_{\rm PNS}-L^{(1)}_{\rm prog}$ at 1~s, requiring that it does not exceed the neutrino luminosity (for our reference model $L_\nu=4.4\times 10^{52}\rm \, erg/s$~\cite{Caputo:2022mah,Fiorillo:2022cdq}).

{\bf\textit{Particle constraints in the trapping regime.}}---We now examine the practical impact of diffusive energy transfer on the constraints for new particles. Fig~\ref{fig:fig2} presents these constraints using our newly derived Eq.~\eqref{eq:e_dep_correct}, compared to those of Eq.~\eqref{eq:wrong_eq_1}. Even in the free-streaming regime, our constraints differ by 10\% from those in Ref.~\cite{Caputo:2022mah} because we do not account for gravitational lapse effects. These effects are significantly smaller than the uncertainties in the SN emission model, which we take to be the Garching group’s muonic model SFHo-18.8~\cite{JankaWeb}. This model, often referred to as the ``cold model'', has been widely used to set limits on new particles, as in Refs.~\cite{Bollig:2020xdr, Caputo:2021rux,Caputo:2022mah, Fiorillo:2022cdq, Fiorillo:2024upk, Telalovic:2024cot}. Moreover, in the trapping regime, a fully consistent treatment of gravitational effects would be considerably more complex, since axion trajectories would be influenced by gravity to an extent comparable to redshift effects. 

For large masses, Eqs.~\eqref{eq:wrong_eq_1} and~\eqref{eq:e_dep_correct} coincide, since the axion mass is so large that even the peak energy transfer in Fig.~\ref{fig:fig1} reaches the $1\,\rm B$ threshold. The constraints are dominated by the non-diffusive energy transfer region---the axiosphere is well within the PNS, and Eq.~\eqref{eq:wrong_eq_1} holds. However, as the mass decreases, the constraints derived from Eq.~\eqref{eq:wrong_eq_1} abruptly extend to infinitely large couplings; even with an infinitesimal mean free path, Eq.~\eqref{eq:wrong_eq_1} predicts a constant energy flux exceeding the limit set by LESNe. It is now clear that this is unphysical and that the energy transport instead enters the diffusive regime, leading to a ceiling in the constraints. In Ref.~\cite{Caputo:2022mah}, for smaller masses, the constraints were extrapolated as a power law. However, Fig.~\ref{fig:fig2} shows that in reality, they deviate from this behavior, shifting trend due to the transition to diffusive heat transport---precisely corresponding to the slope change in Fig.~\ref{fig:fig1} from peak energy transfer to the diffusive transport region. At lower masses, the constraints in the trapping regime can be accurately obtained using the much simpler Eq.~\eqref{eq:diffusive_energy_transfer}.

{\bf\textit{Discussion and outlook.}}---Energy deposition in the material surrounding the proto-neutron star formed in core-collapse supernovae has long been recognized as a powerful probe for feebly interacting particles decaying into photons and charged leptons~\cite{Falk:1978kf,Sung:2019xie,Caputo:2021rux,Calore:2021lih,Caputo:2022mah,Chauhan:2023sci,Chauhan:2024nfa}. For the first time, we have extended these constraints to the trapping regime, imposing bounds based on the requirement that the total explosion energy of normal core-collapse supernovae and low-energy supernovae does not exceed $1 \rm \,B$ and $0.1 \rm\, B$, respectively. Our main conclusion is that previously proposed methods can yield constraints dramatically different from our revised results. Energy deposition in the trapping regime is diffusive, with axions nearly in equilibrium extracting heat from the PNS, and dissipating it via decay at larger radii. A potential concern is that this dissipation might happen within material still infalling on the PNS; thus, we also show a very conservative estimate in SupM~\cite{supplementalmaterial} where only the energy deposited after 300~ms, when the explosion has set in and there is no further accreting material, is considered.


Although we have focused on the case of an axion coupling to two photons, our argument, encapsulated in Eq.~\eqref{eq:e_dep_correct}, applies to any radiatively decaying particle, including sterile neutrinos, novel gauge bosons, and axions coupling to charged leptons. Many of these particles are actively searched for in beam-dump experiments, and low-energy SNe provide a complementary probe by closing the parameter space from below. For instance, our new argument is essential to confirm that regions such as the so-called ``cosmological triangle"~\cite{Brdar:2020dpr}---a region of the axion parameter space previously excluded only by cosmology---are in tension with the energy deposition. For novel gauge bosons, for example, $U(1)_{L_\mu-L_\tau}$ models, our approach could be used to probe parameter regions that are relevant to particle physics and cosmological tensions~\cite{Escudero:2019gzq}. Likewise, our bounds should apply to heavy neutral leptons, see e.g.~\cite{Magill:2018jla}. As an immediate application, we revisit in an upcoming paper the constraints on axion-electron couplings including the new prescription for trapping~\cite{SNElectrons}.

We expect our results to be pertinent also to scalars coupling to leptons or nucleons, which are often considered a viable portal between dark matter and ordinary matter in sub-GeV dark matter searches~(see e.g.~\cite{Essig:2011nj,Hochberg:2015pha,Hochberg:2015fth,Knapen:2016cue,Schutz:2016tid,Knapen:2017ekk,Hochberg:2017wce,Gelmini:2020xir,Hochberg:2021pkt,Knapen:2021run,Knapen:2021bwg,Hochberg:2021yud,CDEX:2022kcd,Boyd:2022tcn,Catena:2022fnk,Gao:2024irf,QROCODILE:2024zmg}, and the Snowmass proceeding~\cite{Essig:2022dfa}). It is possible that the region between the trapping regime of the cooling bound---as treated in the existing literature~\cite{Krnjaic:2015mbs,Knapen:2017xzo,Balaji:2022noj,Hardy:2024gwy}---and laboratory constraints is ruled out by energy deposition up to masses of several hundreds of MeV. If so, the cross section for dark matter interacting via a light mediator would be constrained to very small values. 
We plan to explore these possibilities further in future work.

The criterion we propose generalizes the energy-deposition argument to the region of large couplings, the most essential for the interplay with Earth-based searches, showing that in this regime axions are produced in the proto-neutron star and decay in its immediate surroundings, \textit{and vice versa}, with a net heat flux into the surrounding material. Our argument defines a new state of the art for bounds obtained post-processing supernova models; the final word on the effect that novel particles have on supernova physics can be written only through three-dimensional simulations self-consistently accounting for their existence, which currently seem still unfeasible. The discovery of feebly-interacting particles at beam-dump experiments close to a region excluded by our argument would imply the daunting need to include novel particles in both supernova simulations and in the analysis of supernova light curves.

\acknowledgments

{\bf\textit{Acknowledgments.}}---We thank Yonit Hochberg, Hans-Thomas Janka, and Georg Raffelt for comments on a first draft of the paper.
DFGF is supported by the Alexander von Humboldt Foundation (Germany). TP acknowledges support from NSF grant PHY-2020275 (Network
for Neutrinos, Nuclear Astrophysics, and Symmetries (N3AS)).
EV acknowledges support from the Italian MUR Departments of Excellence grant 2023-2027 ``Quantum Frontiers'' and by Istituto Nazionale di Fisica Nucleare (INFN) through the Theoretical Astroparticle Physics (TAsP) project.

\bibliographystyle{bibi}
\bibliography{References}

@misc{supplementalmaterial,
  note = {See Supplemental Material for a recollection of standard results on  axion production, and for the revised constraints neglecting the energy deposited before the explosion.},
  howpublished = {\url{https://}} 
}

@article{Alda:2024cxn,
    author = "Alda, Jorge and Levati, Gabriele and Paradisi, Paride and Rigolin, Stefano and Selimovic, Nudzeim",
    title = "{Collider and astrophysical signatures of light scalars with enhanced $\tau$ couplings}",
    eprint = "2407.18296",
    archivePrefix = "arXiv",
    primaryClass = "hep-ph",
    month = "7",
    year = "2024"
}

@article{Akita:2024nam,
    author = "Akita, Kensuke and Baur, Gideon and Ovchynnikov, Maksym and Schwetz, Thomas and Syvolap, Vsevolod",
    title = "{New physics decaying into metastable particles: impact on cosmic neutrinos}",
    eprint = "2411.00892",
    archivePrefix = "arXiv",
    primaryClass = "hep-ph",
    reportNumber = "CERN-TH-2024-188, CERN-TH-2024-188",
    month = "10",
    year = "2024"
}

@article{Calore:2020tjw,
    author = "Calore, Francesca and Carenza, Pierluca and Giannotti, Maurizio and Jaeckel, Joerg and Mirizzi, Alessandro",
    title = "{Bounds on axionlike particles from the diffuse supernova flux}",
    eprint = "2008.11741",
    archivePrefix = "arXiv",
    primaryClass = "hep-ph",
    doi = "10.1103/PhysRevD.102.123005",
    journal = "Phys. Rev. D",
    volume = "102",
    number = "12",
    pages = "123005",
    year = "2020"
}

@article{CHARM:1985anb,
    author = "Bergsma, F. and others",
    collaboration = "CHARM",
    title = "{Search for Axion Like Particle Production in 400-{GeV} Proton - Copper Interactions}",
    reportNumber = "CERN-EP-85-38",
    doi = "10.1016/0370-2693(85)90400-9",
    journal = "Phys. Lett. B",
    volume = "157",
    pages = "458--462",
    year = "1985"
}

@article{Abdullahi:2022jlv,
    author = "Abdullahi, Asli M. and others",
    title = "{The present and future status of heavy neutral leptons}",
    eprint = "2203.08039",
    archivePrefix = "arXiv",
    primaryClass = "hep-ph",
    reportNumber = "FERMILAB-CONF-22-184-T-V",
    doi = "10.1088/1361-6471/ac98f9",
    journal = "J. Phys. G",
    volume = "50",
    number = "2",
    pages = "020501",
    year = "2023"
}

@inproceedings{Baryakhtar:2022hbu,
    author = "Baryakhtar, Masha and others",
    title = "{Dark Matter In Extreme Astrophysical Environments}",
    booktitle = "{Snowmass 2021}",
    eprint = "2203.07984",
    archivePrefix = "arXiv",
    primaryClass = "hep-ph",
    month = "3",
    year = "2022"
}

@article{Berger:2016vxi,
    author = "Berger, Joshua and Jedamzik, Karsten and Walker, Devin G. E.",
    title = "{Cosmological Constraints on Decoupled Dark Photons and Dark Higgs}",
    eprint = "1605.07195",
    archivePrefix = "arXiv",
    primaryClass = "hep-ph",
    reportNumber = "SLAC-PUB-16533",
    doi = "10.1088/1475-7516/2016/11/032",
    journal = "JCAP",
    volume = "11",
    pages = "032",
    year = "2016"
}

@article{Li:2025yzb,
    author = "Li, Haotian and Liu, Zuowei and Song, Ningqiang",
    title = "{Probing axion and muon-philic new physics with muon beam dump}",
    eprint = "2501.06294",
    archivePrefix = "arXiv",
    primaryClass = "hep-ph",
    month = "1",
    year = "2025"
}

@article{Boyarsky:2020dzc,
    author = "Boyarsky, Alexey and Ovchynnikov, Maksym and Ruchayskiy, Oleg and Syvolap, Vsevolod",
    title = "{Improved big bang nucleosynthesis constraints on heavy neutral leptons}",
    eprint = "2008.00749",
    archivePrefix = "arXiv",
    primaryClass = "hep-ph",
    doi = "10.1103/PhysRevD.104.023517",
    journal = "Phys. Rev. D",
    volume = "104",
    number = "2",
    pages = "023517",
    year = "2021"
}

@article{Falk:1978kf,
    author = "Falk, Sydney W. and Schramm, David N.",
    title = "{Limits From Supernovae on Neutrino Radiative Lifetimes}",
    reportNumber = "EFI-78-35-CHICAGO",
    doi = "10.1016/0370-2693(78)90417-3",
    journal = "Phys. Lett. B",
    volume = "79",
    pages = "511",
    year = "1978"
}

@article{QROCODILE:2024zmg,
    author = "Baudis, Laura and others",
    collaboration = "QROCODILE",
    title = "{A New Bite Into Dark Matter with the SNSPD-Based QROCODILE Experiment}",
    eprint = "2412.16279",
    archivePrefix = "arXiv",
    primaryClass = "hep-ph",
    reportNumber = "MIT-CTP/5744",
    month = "12",
    year = "2024"
}

@article{Gao:2024irf,
    author = "Gao, Jiansong and Hochberg, Yonit and Lehmann, Benjamin V. and Nam, Sae Woo and Szypryt, Paul and Vissers, Michael R. and Xu, Tao",
    title = "{Detecting Light Dark Matter with Kinetic Inductance Detectors}",
    eprint = "2403.19739",
    archivePrefix = "arXiv",
    primaryClass = "hep-ph",
    reportNumber = "MIT-CTP/5654",
    month = "3",
    year = "2024"
}

@article{Hochberg:2021yud,
    author = "Hochberg, Yonit and Lehmann, Benjamin V. and Charaev, Ilya and Chiles, Jeff and Colangelo, Marco and Nam, Sae Woo and Berggren, Karl K.",
    title = "{New constraints on dark matter from superconducting nanowires}",
    eprint = "2110.01586",
    archivePrefix = "arXiv",
    primaryClass = "hep-ph",
    doi = "10.1103/PhysRevD.106.112005",
    journal = "Phys. Rev. D",
    volume = "106",
    number = "11",
    pages = "112005",
    year = "2022"
}

@article{Candon:2024eah,
    author = "Cand\'on, Francisco R. and Fiorillo, Damiano F. G. and Lucente, Giuseppe and Vitagliano, Edoardo and Vogel, Julia K.",
    title = "{NuSTAR bounds on radiatively decaying particles from M82}",
    eprint = "2412.03660",
    archivePrefix = "arXiv",
    primaryClass = "hep-ph",
    month = "12",
    year = "2024"
}

@article{Jaeckel:2017tud,
    author = "Jaeckel, J. and Malta, P. C. and Redondo, J.",
    title = "{Decay photons from the axionlike particles burst of type II supernovae}",
    eprint = "1702.02964",
    archivePrefix = "arXiv",
    primaryClass = "hep-ph",
    doi = "10.1103/PhysRevD.98.055032",
    journal = "Phys. Rev. D",
    volume = "98",
    number = "5",
    pages = "055032",
    year = "2018"
}

@article{Chauhan:2023sci,
    author = "Chauhan, Garv and Horiuchi, Shunsaku and Huber, Patrick and Shoemaker, Ian M.",
    title = "{Low-Energy Supernovae Bounds on Sterile Neutrinos}",
    eprint = "2309.05860",
    archivePrefix = "arXiv",
    primaryClass = "hep-ph",
    month = "9",
    year = "2023"
}

@article{Hardy:2024gwy,
    author = "Hardy, Edward and Sokolov, Anton and Stubbs, Henry",
    title = "{Supernova bounds on new scalars from resonant and soft emission}",
    eprint = "2410.17347",
    archivePrefix = "arXiv",
    primaryClass = "hep-ph",
    doi = "10.1007/JHEP04(2025)013",
    journal = "JHEP",
    volume = "04",
    pages = "013",
    year = "2025"
}

@article{Balaji:2022noj,
    author = "Balaji, Shyam and Dev, P. S. Bhupal and Silk, Joseph and Zhang, Yongchao",
    title = "{Improved stellar limits on a light CP-even scalar}",
    eprint = "2205.01669",
    archivePrefix = "arXiv",
    primaryClass = "hep-ph",
    doi = "10.1088/1475-7516/2022/12/024",
    journal = "JCAP",
    volume = "12",
    pages = "024",
    year = "2022"
}

@article{Knapen:2021bwg,
    author = "Knapen, Simon and Kozaczuk, Jonathan and Lin, Tongyan",
    title = "{python package for dark matter scattering in dielectric targets}",
    eprint = "2104.12786",
    archivePrefix = "arXiv",
    primaryClass = "hep-ph",
    doi = "10.1103/PhysRevD.105.015014",
    journal = "Phys. Rev. D",
    volume = "105",
    number = "1",
    pages = "015014",
    year = "2022"
}

@inproceedings{Essig:2022dfa,
    author = "Essig, Rouven and others",
    title = "{Snowmass2021 Cosmic Frontier: The landscape of low-threshold dark matter direct detection in the next decade}",
    booktitle = "{Snowmass 2021}",
    eprint = "2203.08297",
    archivePrefix = "arXiv",
    primaryClass = "hep-ph",
    reportNumber = "FERMILAB-CONF-22-181-PPD",
    month = "3",
    year = "2022"
}

@article{Mayle:1987as,
    author = "Mayle, Ron and Wilson, James R. and Ellis, John R. and Olive, Keith A. and Schramm, David N. and Steigman, Gary",
    title = "{Constraints on Axions from SN 1987a}",
    reportNumber = "FERMILAB-PUB-87-225-A, EFI-87-104-CHICAGO, UMN-TH-637-87, CERN-TH-4887-87",
    doi = "10.1016/0370-2693(88)91595-X",
    journal = "Phys. Lett. B",
    volume = "203",
    pages = "188--196",
    year = "1988"
}

@article{Mayle:1989yx,
    author = "Mayle, Ron and Wilson, James R. and Ellis, John R. and Olive, Keith A. and Schramm, David N. and Steigman, Gary",
    title = "{Updated Constraints on Axions from SN 1987a}",
    reportNumber = "PRINT-89-0012 (OHIO-STATE), FERMILAB-PUB-88-209-A",
    doi = "10.1016/0370-2693(89)91104-0",
    journal = "Phys. Lett. B",
    volume = "219",
    pages = "515",
    year = "1989"
}

@article{Turner:1987by,
    author = "Turner, Michael S.",
    title = "{Axions from SN 1987a}",
    reportNumber = "FERMILAB-PUB-87-202-A",
    doi = "10.1103/PhysRevLett.60.1797",
    journal = "Phys. Rev. Lett.",
    volume = "60",
    pages = "1797",
    year = "1988"
}

@article{Burrows:1990pk,
    author = "Burrows, Adam and Ressell, M. Ted and Turner, Michael S.",
    title = "{Axions and SN1987A: Axion trapping}",
    reportNumber = "FERMILAB-PUB-90-081-A, FERMILAB-PUB-90-081-A-REV",
    doi = "10.1103/PhysRevD.42.3297",
    journal = "Phys. Rev. D",
    volume = "42",
    pages = "3297--3309",
    year = "1990"
}

@article{Burrows:1988ah,
    author = "Burrows, Adam and Turner, Michael S. and Brinkmann, R. P.",
    title = "{Axions and SN 1987a}",
    reportNumber = "FERMILAB-PUB-88-105-A",
    doi = "10.1103/PhysRevD.39.1020",
    journal = "Phys. Rev. D",
    volume = "39",
    pages = "1020",
    year = "1989"
}

@book{Raffelt:1996wa,
    author = "Raffelt, G. G.",
    title = "{Stars as laboratories for fundamental physics}: {The astrophysics of neutrinos, axions, and other weakly interacting particles}",
    isbn = "978-0-226-70272-8",
    month = "5",
    year = "1996"
}

@inproceedings{Essig:2013lka,
    author = "Essig, Rouven and others",
    title = "{Working Group Report: New Light Weakly Coupled Particles}",
    booktitle = "{Snowmass 2013}: {Snowmass on the Mississippi}",
    eprint = "1311.0029",
    archivePrefix = "arXiv",
    primaryClass = "hep-ph",
    reportNumber = "YITP-SB-36, FERMILAB-CONF-13-653",
    month = "10",
    year = "2013"
}

@article{Hochberg:2015fth,
    author = "Hochberg, Yonit and Pyle, Matt and Zhao, Yue and Zurek, Kathryn M.",
    title = "{Detecting Superlight Dark Matter with Fermi-Degenerate Materials}",
    eprint = "1512.04533",
    archivePrefix = "arXiv",
    primaryClass = "hep-ph",
    doi = "10.1007/JHEP08(2016)057",
    journal = "JHEP",
    volume = "08",
    pages = "057",
    year = "2016"
}

@article{Hochberg:2015pha,
    author = "Hochberg, Yonit and Zhao, Yue and Zurek, Kathryn M.",
    title = "{Superconducting Detectors for Superlight Dark Matter}",
    eprint = "1504.07237",
    archivePrefix = "arXiv",
    primaryClass = "hep-ph",
    doi = "10.1103/PhysRevLett.116.011301",
    journal = "Phys. Rev. Lett.",
    volume = "116",
    number = "1",
    pages = "011301",
    year = "2016"
}

@article{Knapen:2016cue,
    author = "Knapen, Simon and Lin, Tongyan and Zurek, Kathryn M.",
    title = "{Light Dark Matter in Superfluid Helium: Detection with Multi-excitation Production}",
    eprint = "1611.06228",
    archivePrefix = "arXiv",
    primaryClass = "hep-ph",
    doi = "10.1103/PhysRevD.95.056019",
    journal = "Phys. Rev. D",
    volume = "95",
    number = "5",
    pages = "056019",
    year = "2017"
}

@article{Schutz:2016tid,
    author = "Schutz, Katelin and Zurek, Kathryn M.",
    title = "{Detectability of Light Dark Matter with Superfluid Helium}",
    eprint = "1604.08206",
    archivePrefix = "arXiv",
    primaryClass = "hep-ph",
    doi = "10.1103/PhysRevLett.117.121302",
    journal = "Phys. Rev. Lett.",
    volume = "117",
    number = "12",
    pages = "121302",
    year = "2016"
}

@article{Knapen:2017ekk,
    author = "Knapen, Simon and Lin, Tongyan and Pyle, Matt and Zurek, Kathryn M.",
    title = "{Detection of Light Dark Matter With Optical Phonons in Polar Materials}",
    eprint = "1712.06598",
    archivePrefix = "arXiv",
    primaryClass = "hep-ph",
    doi = "10.1016/j.physletb.2018.08.064",
    journal = "Phys. Lett. B",
    volume = "785",
    pages = "386--390",
    year = "2018"
}

@article{Catena:2022fnk,
    author = "Catena, Riccardo and Cole, Daniel and Emken, Timon and Matas, Marek and Spaldin, Nicola and Tarantino, Walter and Urdshals, Einar",
    title = "{Dark matter-electron interactions in materials beyond the dark photon model}",
    eprint = "2210.07305",
    archivePrefix = "arXiv",
    primaryClass = "hep-ph",
    doi = "10.1088/1475-7516/2023/03/052",
    journal = "JCAP",
    volume = "03",
    pages = "052",
    year = "2023"
}

@article{CDEX:2022kcd,
    author = "Zhang, Z. Y. and others",
    collaboration = "CDEX",
    title = "{Constraints on Sub-GeV Dark Matter\textendash{}Electron Scattering from the CDEX-10 Experiment}",
    eprint = "2206.04128",
    archivePrefix = "arXiv",
    primaryClass = "hep-ex",
    doi = "10.1103/PhysRevLett.129.221301",
    journal = "Phys. Rev. Lett.",
    volume = "129",
    number = "22",
    pages = "221301",
    year = "2022"
}

@article{Knapen:2021run,
    author = "Knapen, Simon and Kozaczuk, Jonathan and Lin, Tongyan",
    title = "{Dark matter-electron scattering in dielectrics}",
    eprint = "2101.08275",
    archivePrefix = "arXiv",
    primaryClass = "hep-ph",
    reportNumber = "CERN-TH-2021-013",
    doi = "10.1103/PhysRevD.104.015031",
    journal = "Phys. Rev. D",
    volume = "104",
    number = "1",
    pages = "015031",
    year = "2021"
}

@article{Boyd:2022tcn,
    author = "Boyd, Christian and Hochberg, Yonit and Kahn, Yonatan and Kramer, Eric David and Kurinsky, Noah and Lehmann, Benjamin V. and Yu, To Chin",
    title = "{Directional detection of dark matter with anisotropic response functions}",
    eprint = "2212.04505",
    archivePrefix = "arXiv",
    primaryClass = "hep-ph",
    doi = "10.1103/PhysRevD.108.015015",
    journal = "Phys. Rev. D",
    volume = "108",
    number = "1",
    pages = "015015",
    year = "2023"
}

@article{Hochberg:2021pkt,
    author = "Hochberg, Yonit and Kahn, Yonatan and Kurinsky, Noah and Lehmann, Benjamin V. and Yu, To Chin and Berggren, Karl K.",
    title = "{Determining Dark-Matter\textendash{}Electron Scattering Rates from the Dielectric Function}",
    eprint = "2101.08263",
    archivePrefix = "arXiv",
    primaryClass = "hep-ph",
    reportNumber = "FERMILAB-PUB-21-128-AE",
    doi = "10.1103/PhysRevLett.127.151802",
    journal = "Phys. Rev. Lett.",
    volume = "127",
    number = "15",
    pages = "151802",
    year = "2021"
}

@article{Gelmini:2020xir,
    author = "Gelmini, Graciela B. and Takhistov, Volodymyr and Vitagliano, Edoardo",
    title = "{Scalar direct detection: In-medium effects}",
    eprint = "2006.13909",
    archivePrefix = "arXiv",
    primaryClass = "hep-ph",
    doi = "10.1016/j.physletb.2020.135779",
    journal = "Phys. Lett. B",
    volume = "809",
    pages = "135779",
    year = "2020"
}

@article{Hochberg:2017wce,
    author = "Hochberg, Yonit and Kahn, Yonatan and Lisanti, Mariangela and Zurek, Kathryn M. and Grushin, Adolfo G. and Ilan, Roni and Griffin, Sin\'ead M. and Liu, Zhen-Fei and Weber, Sophie F. and Neaton, Jeffrey B.",
    title = "{Detection of sub-MeV Dark Matter with Three-Dimensional Dirac Materials}",
    eprint = "1708.08929",
    archivePrefix = "arXiv",
    primaryClass = "hep-ph",
    reportNumber = "PUPT-2535",
    doi = "10.1103/PhysRevD.97.015004",
    journal = "Phys. Rev. D",
    volume = "97",
    number = "1",
    pages = "015004",
    year = "2018"
}

@article{Essig:2011nj,
    author = "Essig, Rouven and Mardon, Jeremy and Volansky, Tomer",
    title = "{Direct Detection of Sub-GeV Dark Matter}",
    eprint = "1108.5383",
    archivePrefix = "arXiv",
    primaryClass = "hep-ph",
    reportNumber = "SLAC-PUB-14538",
    doi = "10.1103/PhysRevD.85.076007",
    journal = "Phys. Rev. D",
    volume = "85",
    pages = "076007",
    year = "2012"
}

@article{Krnjaic:2015mbs,
    author = "Krnjaic, Gordan",
    title = "{Probing Light Thermal Dark-Matter With a Higgs Portal Mediator}",
    eprint = "1512.04119",
    archivePrefix = "arXiv",
    primaryClass = "hep-ph",
    reportNumber = "FERMILAB-PUB-15-550-A",
    doi = "10.1103/PhysRevD.94.073009",
    journal = "Phys. Rev. D",
    volume = "94",
    number = "7",
    pages = "073009",
    year = "2016"
}

@article{Knapen:2017xzo,
    author = "Knapen, Simon and Lin, Tongyan and Zurek, Kathryn M.",
    title = "{Light Dark Matter: Models and Constraints}",
    eprint = "1709.07882",
    archivePrefix = "arXiv",
    primaryClass = "hep-ph",
    doi = "10.1103/PhysRevD.96.115021",
    journal = "Phys. Rev. D",
    volume = "96",
    number = "11",
    pages = "115021",
    year = "2017"
}

@article{Brdar:2020dpr,
    author = "Brdar, Vedran and Dutta, Bhaskar and Jang, Wooyoung and Kim, Doojin and Shoemaker, Ian M. and Tabrizi, Zahra and Thompson, Adrian and Yu, Jaehoon",
    title = "{Axionlike Particles at Future Neutrino Experiments: Closing the Cosmological Triangle}",
    eprint = "2011.07054",
    archivePrefix = "arXiv",
    primaryClass = "hep-ph",
    reportNumber = "FERMILAB-PUB-20-645-V, MI-TH-2029",
    doi = "10.1103/PhysRevLett.126.201801",
    journal = "Phys. Rev. Lett.",
    volume = "126",
    number = "20",
    pages = "201801",
    year = "2021"
}

@article{Calore:2021lih,
    author = "Calore, Francesca and Carenza, Pierluca and Giannotti, Maurizio and Jaeckel, Joerg and Lucente, Giuseppe and Mastrototaro, Leonardo and Mirizzi, Alessandro",
    title = "{511~keV line constraints on feebly interacting particles from supernovae}",
    eprint = "2112.08382",
    archivePrefix = "arXiv",
    primaryClass = "hep-ph",
    doi = "10.1103/PhysRevD.105.063026",
    journal = "Phys. Rev. D",
    volume = "105",
    number = "6",
    pages = "063026",
    year = "2022"
}

@article{Chauhan:2024nfa,
    author = "Chauhan, Garv and Horiuchi, Shunsaku and Huber, Patrick and Shoemaker, Ian M.",
    title = "{Probing the sterile neutrino dipole portal with SN1987A and low-energy supernovae}",
    eprint = "2402.01624",
    archivePrefix = "arXiv",
    primaryClass = "hep-ph",
    doi = "10.1103/PhysRevD.110.015007",
    journal = "Phys. Rev. D",
    volume = "110",
    number = "1",
    pages = "015007",
    year = "2024"
}

@article{Fiorillo:2023frv,
    author = "Fiorillo, Damiano F. G. and Heinlein, Malte and Janka, Hans-Thomas and Raffelt, Georg and Vitagliano, Edoardo and Bollig, Robert",
    title = "{Supernova simulations confront SN 1987A neutrinos}",
    eprint = "2308.01403",
    archivePrefix = "arXiv",
    primaryClass = "astro-ph.HE",
    doi = "10.1103/PhysRevD.108.083040",
    journal = "Phys. Rev. D",
    volume = "108",
    number = "8",
    pages = "083040",
    year = "2023"
}

@article{Sung:2019xie,
    author = "Sung, Allan and Tu, Huitzu and Wu, Meng-Ru",
    title = "{New constraint from supernova explosions on light particles beyond the Standard Model}",
    eprint = "1903.07923",
    archivePrefix = "arXiv",
    primaryClass = "hep-ph",
    doi = "10.1103/PhysRevD.99.121305",
    journal = "Phys. Rev. D",
    volume = "99",
    number = "12",
    pages = "121305",
    year = "2019"
}

@article{Magill:2018jla,
    author = "Magill, Gabriel and Plestid, Ryan and Pospelov, Maxim and Tsai, Yu-Dai",
    title = "{Dipole Portal to Heavy Neutral Leptons}",
    eprint = "1803.03262",
    archivePrefix = "arXiv",
    primaryClass = "hep-ph",
    reportNumber = "FERMILAB-PUB-18-745-A",
    doi = "10.1103/PhysRevD.98.115015",
    journal = "Phys. Rev. D",
    volume = "98",
    number = "11",
    pages = "115015",
    year = "2018"
}

@article{Bolton:2019pcu,
    author = "Bolton, Patrick D. and Deppisch, Frank F. and Bhupal Dev, P. S.",
    title = "{Neutrinoless double beta decay versus other probes of heavy sterile neutrinos}",
    eprint = "1912.03058",
    archivePrefix = "arXiv",
    primaryClass = "hep-ph",
    doi = "10.1007/JHEP03(2020)170",
    journal = "JHEP",
    volume = "03",
    pages = "170",
    year = "2020"
}

@article{Sabti:2020yrt,
    author = "Sabti, Nashwan and Magalich, Andrii and Filimonova, Anastasiia",
    title = "{An Extended Analysis of Heavy Neutral Leptons during Big Bang Nucleosynthesis}",
    eprint = "2006.07387",
    archivePrefix = "arXiv",
    primaryClass = "hep-ph",
    reportNumber = "KCL-2020-09",
    doi = "10.1088/1475-7516/2020/11/056",
    journal = "JCAP",
    volume = "11",
    pages = "056",
    year = "2020"
}

@article{Mastrototaro:2021wzl,
    author = "Mastrototaro, Leonardo and Serpico, Pasquale Dario and Mirizzi, Alessandro and Saviano, Ninetta",
    title = "{Massive sterile neutrinos in the early Universe: From thermal decoupling to cosmological constraints}",
    eprint = "2104.11752",
    archivePrefix = "arXiv",
    primaryClass = "hep-ph",
    reportNumber = "LAPTH-017/21",
    doi = "10.1103/PhysRevD.104.016026",
    journal = "Phys. Rev. D",
    volume = "104",
    number = "1",
    pages = "016026",
    year = "2021"
}

@article{Fradette:2014sza,
    author = "Fradette, Anthony and Pospelov, Maxim and Pradler, Josef and Ritz, Adam",
    title = "{Cosmological Constraints on Very Dark Photons}",
    eprint = "1407.0993",
    archivePrefix = "arXiv",
    primaryClass = "hep-ph",
    doi = "10.1103/PhysRevD.90.035022",
    journal = "Phys. Rev. D",
    volume = "90",
    number = "3",
    pages = "035022",
    year = "2014"
}

@article{Redondo:2008ec,
    author = "Redondo, Javier and Postma, Marieke",
    title = "{Massive hidden photons as lukewarm dark matter}",
    eprint = "0811.0326",
    archivePrefix = "arXiv",
    primaryClass = "hep-ph",
    reportNumber = "NIKHEF-2008-030, DESY-08-154",
    doi = "10.1088/1475-7516/2009/02/005",
    journal = "JCAP",
    volume = "02",
    pages = "005",
    year = "2009"
}

@article{Fradette:2018hhl,
    author = "Fradette, Anthony and Pospelov, Maxim and Pradler, Josef and Ritz, Adam",
    title = "{Cosmological beam dump: constraints on dark scalars mixed with the Higgs boson}",
    eprint = "1812.07585",
    archivePrefix = "arXiv",
    primaryClass = "hep-ph",
    doi = "10.1103/PhysRevD.99.075004",
    journal = "Phys. Rev. D",
    volume = "99",
    number = "7",
    pages = "075004",
    year = "2019"
}

@article{DEramo:2024lsk,
    author = "D'Eramo, Francesco and Tesi, Andrea and Vaskonen, Ville",
    title = "{Irreducible cosmological backgrounds of a real scalar with a broken symmetry}",
    eprint = "2407.19997",
    archivePrefix = "arXiv",
    primaryClass = "hep-ph",
    doi = "10.1103/PhysRevD.110.095002",
    journal = "Phys. Rev. D",
    volume = "110",
    number = "9",
    pages = "095002",
    year = "2024"
}

@article{Escudero:2019gzq,
    author = "Escudero, Miguel and Hooper, Dan and Krnjaic, Gordan and Pierre, Mathias",
    title = "{Cosmology with A Very Light L$_{\mu}$ \ensuremath{-} L$_{\tau}$ Gauge Boson}",
    eprint = "1901.02010",
    archivePrefix = "arXiv",
    primaryClass = "hep-ph",
    reportNumber = "FERMILAB-PUB-19-001-A, LPT-Orsay-18-15, IFIC-19-02, KCL-19-01,
  IFT-UAM/CSIC-19-7, KCL-19-01",
    doi = "10.1007/JHEP03(2019)071",
    journal = "JHEP",
    volume = "03",
    pages = "071",
    year = "2019"
}

@article{Langhoff:2022bij,
    author = "Langhoff, Kevin and Outmezguine, Nadav Joseph and Rodd, Nicholas L.",
    title = "{Irreducible Axion Background}",
    eprint = "2209.06216",
    archivePrefix = "arXiv",
    primaryClass = "hep-ph",
    reportNumber = "CERN-TH-2022-148",
    doi = "10.1103/PhysRevLett.129.241101",
    journal = "Phys. Rev. Lett.",
    volume = "129",
    number = "24",
    pages = "241101",
    year = "2022"
}

@article{Depta:2020zbh,
    author = "Depta, Paul Frederik and Hufnagel, Marco and Schmidt-Hoberg, Kai",
    title = "{Updated BBN constraints on electromagnetic decays of MeV-scale particles}",
    eprint = "2011.06519",
    archivePrefix = "arXiv",
    primaryClass = "hep-ph",
    reportNumber = "DESY-20-160, DESY 20-160, ULB-TH/20-15",
    doi = "10.1088/1475-7516/2021/04/011",
    journal = "JCAP",
    volume = "04",
    pages = "011",
    year = "2021"
}

@article{Depta:2019lbe,
    author = "Depta, Paul Frederik and Hufnagel, Marco and Schmidt-Hoberg, Kai and Wild, Sebastian",
    title = "{BBN constraints on the annihilation of MeV-scale dark matter}",
    eprint = "1901.06944",
    archivePrefix = "arXiv",
    primaryClass = "hep-ph",
    reportNumber = "DESY-19-006, DESY 19-006",
    doi = "10.1088/1475-7516/2019/04/029",
    journal = "JCAP",
    volume = "04",
    pages = "029",
    year = "2019"
}

@article{Depta:2020wmr,
    author = "Depta, Paul Frederik and Hufnagel, Marco and Schmidt-Hoberg, Kai",
    title = "{Robust cosmological constraints on axion-like particles}",
    eprint = "2002.08370",
    archivePrefix = "arXiv",
    primaryClass = "hep-ph",
    reportNumber = "DESY-20-003, DESY 20-003",
    doi = "10.1088/1475-7516/2020/05/009",
    journal = "JCAP",
    volume = "05",
    pages = "009",
    year = "2020"
}

@article{Cadamuro:2011fd,
    author = "Cadamuro, Davide and Redondo, Javier",
    title = "{Cosmological bounds on pseudo Nambu-Goldstone bosons}",
    eprint = "1110.2895",
    archivePrefix = "arXiv",
    primaryClass = "hep-ph",
    reportNumber = "MPP-2011-116",
    doi = "10.1088/1475-7516/2012/02/032",
    journal = "JCAP",
    volume = "02",
    pages = "032",
    year = "2012"
}

@article{Agrawal:2021dbo,
    author = "Agrawal, Prateek and others",
    title = "{Feebly-interacting particles: FIPs 2020 workshop report}",
    eprint = "2102.12143",
    archivePrefix = "arXiv",
    primaryClass = "hep-ph",
    doi = "10.1140/epjc/s10052-021-09703-7",
    journal = "Eur. Phys. J. C",
    volume = "81",
    number = "11",
    pages = "1015",
    year = "2021"
}

@article{Berryman:2022hds,
    author = "Berryman, Jeffrey M. and others",
    title = "{Neutrino self-interactions: A white paper}",
    eprint = "2203.01955",
    archivePrefix = "arXiv",
    primaryClass = "hep-ph",
    reportNumber = "CERN-TH-2022-024, DESY-22-035, FERMILAB-PUB-22-099-T",
    doi = "10.1016/j.dark.2023.101267",
    journal = "Phys. Dark Univ.",
    volume = "42",
    pages = "101267",
    year = "2023"
}

@article{Knapen:2024fvh,
    author = "Knapen, Simon and Opferkuch, Toby and Redigolo, Diego and Tammaro, Michele",
    title = "{Displaced Searches for Axion-Like Particles and Heavy Neutral Leptons at Mu3e}",
    eprint = "2410.13941",
    archivePrefix = "arXiv",
    primaryClass = "hep-ph",
    month = "10",
    year = "2024"
}

@article{Krnjaic:2019rsv,
    author = "Krnjaic, Gordan and Marques-Tavares, Gustavo and Redigolo, Diego and Tobioka, Kohsaku",
    title = "{Probing Muonphilic Force Carriers and Dark Matter at Kaon Factories}",
    eprint = "1902.07715",
    archivePrefix = "arXiv",
    primaryClass = "hep-ph",
    reportNumber = "FERMILAB-PUB-18-665-A, KEK-TH-2105",
    doi = "10.1103/PhysRevLett.124.041802",
    journal = "Phys. Rev. Lett.",
    volume = "124",
    number = "4",
    pages = "041802",
    year = "2020"
}

@article{Calibbi:2020jvd,
    author = "Calibbi, Lorenzo and Redigolo, Diego and Ziegler, Robert and Zupan, Jure",
    title = "{Looking forward to lepton-flavor-violating ALPs}",
    eprint = "2006.04795",
    archivePrefix = "arXiv",
    primaryClass = "hep-ph",
    reportNumber = "P3H-20-024, TTP20-025",
    doi = "10.1007/JHEP09(2021)173",
    journal = "JHEP",
    volume = "09",
    pages = "173",
    year = "2021"
}

@article{Bauer:2017ris,
    author = "Bauer, Martin and Neubert, Matthias and Thamm, Andrea",
    title = "{Collider Probes of Axion-Like Particles}",
    eprint = "1708.00443",
    archivePrefix = "arXiv",
    primaryClass = "hep-ph",
    reportNumber = "MITP-17-047",
    doi = "10.1007/JHEP12(2017)044",
    journal = "JHEP",
    volume = "12",
    pages = "044",
    year = "2017"
}

@article{Bauer:2018onh,
    author = "Bauer, Martin and Foldenauer, Patrick and Jaeckel, Joerg",
    title = "{Hunting All the Hidden Photons}",
    eprint = "1803.05466",
    archivePrefix = "arXiv",
    primaryClass = "hep-ph",
    doi = "10.1007/JHEP07(2018)094",
    journal = "JHEP",
    volume = "07",
    pages = "094",
    year = "2018"
}

@article{Riordan:1987aw,
    author = "Riordan, E. M. and others",
    title = "{A Search for Short Lived Axions in an Electron Beam Dump Experiment}",
    reportNumber = "SLAC-PUB-4280, UR-993, FERMILAB-PUB-87-251",
    doi = "10.1103/PhysRevLett.59.755",
    journal = "Phys. Rev. Lett.",
    volume = "59",
    pages = "755",
    year = "1987"
}

@article{Blumlein:1990ay,
    author = "Blumlein, J. and others",
    title = "{Limits on neutral light scalar and pseudoscalar particles in a proton beam dump experiment}",
    reportNumber = "PHE-90-03",
    doi = "10.1007/BF01548556",
    journal = "Z. Phys. C",
    volume = "51",
    pages = "341--350",
    year = "1991"
}

@article{NA64:2020qwq,
    author = "Banerjee, D. and others",
    collaboration = "NA64",
    title = "{Search for Axionlike and Scalar Particles with the NA64 Experiment}",
    eprint = "2005.02710",
    archivePrefix = "arXiv",
    primaryClass = "hep-ex",
    reportNumber = "CERN-EP-2020-068",
    doi = "10.1103/PhysRevLett.125.081801",
    journal = "Phys. Rev. Lett.",
    volume = "125",
    number = "8",
    pages = "081801",
    year = "2020"
}

@article{Dolan:2017osp,
    author = "Dolan, Matthew J. and Ferber, Torben and Hearty, Christopher and Kahlhoefer, Felix and Schmidt-Hoberg, Kai",
    title = "{Revised constraints and Belle II sensitivity for visible and invisible axion-like particles}",
    eprint = "1709.00009",
    archivePrefix = "arXiv",
    primaryClass = "hep-ph",
    reportNumber = "DESY-17-127",
    doi = "10.1007/JHEP12(2017)094",
    journal = "JHEP",
    volume = "12",
    pages = "094",
    year = "2017",
    note = "[Erratum: JHEP 03, 190 (2021)]"
}

@article{Capozzi:2023ffu,
    author = "Capozzi, Francesco and Dutta, Bhaskar and Gurung, Gajendra and Jang, Wooyoung and Shoemaker, Ian M. and Thompson, Adrian and Yu, Jaehoon",
    title = "{New constraints on ALP couplings to electrons and photons from ArgoNeuT and the MiniBooNE beam dump}",
    eprint = "2307.03878",
    archivePrefix = "arXiv",
    primaryClass = "hep-ph",
    reportNumber = "MI-HET-808",
    doi = "10.1103/PhysRevD.108.075019",
    journal = "Phys. Rev. D",
    volume = "108",
    number = "7",
    pages = "075019",
    year = "2023"
}

@article{Dev:2023hax,
    author = "Dev, P. S. Bhupal and Fortin, Jean-Fran\c{c}ois and Harris, Steven P. and Sinha, Kuver and Zhang, Yongchao",
    title = "{First Constraints on the Photon Coupling of Axionlike Particles from Multimessenger Studies of the Neutron Star Merger GW170817}",
    eprint = "2305.01002",
    archivePrefix = "arXiv",
    primaryClass = "hep-ph",
    reportNumber = "INT-PUB-23-014",
    doi = "10.1103/PhysRevLett.132.101003",
    journal = "Phys. Rev. Lett.",
    volume = "132",
    number = "10",
    pages = "101003",
    year = "2024"
}

@article{Chang:2018rso,
    author = "Chang, Jae Hyeok and Essig, Rouven and McDermott, Samuel D.",
    title = "{Supernova 1987A Constraints on Sub-GeV Dark Sectors, Millicharged Particles, the QCD Axion, and an Axion-like Particle}",
    eprint = "1803.00993",
    archivePrefix = "arXiv",
    primaryClass = "hep-ph",
    reportNumber = "YITP-SB-18-01, FERMILAB-PUB-17-432-T",
    doi = "10.1007/JHEP09(2018)051",
    journal = "JHEP",
    volume = "09",
    pages = "051",
    year = "2018"
}

@article{Chang:2016ntp,
    author = "Chang, Jae Hyeok and Essig, Rouven and McDermott, Samuel D.",
    title = "{Revisiting Supernova 1987A Constraints on Dark Photons}",
    eprint = "1611.03864",
    archivePrefix = "arXiv",
    primaryClass = "hep-ph",
    reportNumber = "YITP-SB-16-44",
    doi = "10.1007/JHEP01(2017)107",
    journal = "JHEP",
    volume = "01",
    pages = "107",
    year = "2017"
}

@article{Croon:2020lrf,
    author = "Croon, Djuna and Elor, Gilly and Leane, Rebecca K. and McDermott, Samuel D.",
    title = "{Supernova Muons: New Constraints on $Z$' Bosons, Axions and ALPs}",
    eprint = "2006.13942",
    archivePrefix = "arXiv",
    primaryClass = "hep-ph",
    reportNumber = "MIT-CTP/5214, FERMILAB-PUB-20-246-A-T",
    doi = "10.1007/JHEP01(2021)107",
    journal = "JHEP",
    volume = "01",
    pages = "107",
    year = "2021"
}

@article{Camalich:2020wac,
    author = "Camalich, Jorge Martin and Terol-Calvo, Jorge and Tolos, Laura and Ziegler, Robert",
    title = "{Supernova Constraints on Dark Flavored Sectors}",
    eprint = "2012.11632",
    archivePrefix = "arXiv",
    primaryClass = "hep-ph",
    doi = "10.1103/PhysRevD.103.L121301",
    journal = "Phys. Rev. D",
    volume = "103",
    number = "12",
    pages = "L121301",
    year = "2021"
}

@article{Ferreira:2022xlw,
    author = {Ferreira, Ricardo Z. and Marsh, M. C. David and M\"uller, Eike},
    title = "{Strong supernovae bounds on ALPs from quantum loops}",
    eprint = "2205.07896",
    archivePrefix = "arXiv",
    primaryClass = "hep-ph",
    doi = "10.1088/1475-7516/2022/11/057",
    journal = "JCAP",
    volume = "11",
    pages = "057",
    year = "2022"
}

@article{Hoof:2022xbe,
    author = "Hoof, Sebastian and Schulz, Lena",
    title = "{Updated constraints on axion-like particles from temporal information in supernova SN1987A gamma-ray data}",
    eprint = "2212.09764",
    archivePrefix = "arXiv",
    primaryClass = "hep-ph",
    reportNumber = "TTP22-072",
    doi = "10.1088/1475-7516/2023/03/054",
    journal = "JCAP",
    volume = "03",
    pages = "054",
    year = "2023"
}

@article{Lella:2023bfb,
    author = "Lella, Alessandro and Carenza, Pierluca and Co', Giampaolo and Lucente, Giuseppe and Giannotti, Maurizio and Mirizzi, Alessandro and Rauscher, Thomas",
    title = "{Getting the most on supernova axions}",
    eprint = "2306.01048",
    archivePrefix = "arXiv",
    primaryClass = "hep-ph",
    doi = "10.1103/PhysRevD.109.023001",
    journal = "Phys. Rev. D",
    volume = "109",
    number = "2",
    pages = "023001",
    year = "2024"
}

@article{Fiorillo:2024upk,
    author = "Fiorillo, Damiano F. G. and Vitagliano, Edoardo",
    title = "{Self-Interacting Dark Sectors in Supernovae Can Behave as a Relativistic Fluid}",
    eprint = "2404.07714",
    archivePrefix = "arXiv",
    primaryClass = "hep-ph",
    doi = "10.1103/PhysRevLett.133.251004",
    journal = "Phys. Rev. Lett.",
    volume = "133",
    number = "25",
    pages = "251004",
    year = "2024"
}

@article{Diamond:2023scc,
    author = "Diamond, Melissa and Fiorillo, Damiano F. G. and Marques-Tavares, Gustavo and Vitagliano, Edoardo",
    title = "{Axion-sourced fireballs from supernovae}",
    eprint = "2303.11395",
    archivePrefix = "arXiv",
    primaryClass = "hep-ph",
    doi = "10.1103/PhysRevD.107.103029",
    journal = "Phys. Rev. D",
    volume = "107",
    number = "10",
    pages = "103029",
    year = "2023",
    note = "[Erratum: Phys.Rev.D 108, 049902 (2023)]"
}

@article{Fiorillo:2022cdq,
    author = "Fiorillo, Damiano F. G. and Raffelt, Georg G. and Vitagliano, Edoardo",
    title = "{Strong Supernova 1987A Constraints on Bosons Decaying to Neutrinos}",
    eprint = "2209.11773",
    archivePrefix = "arXiv",
    primaryClass = "hep-ph",
    doi = "10.1103/PhysRevLett.131.021001",
    journal = "Phys. Rev. Lett.",
    volume = "131",
    number = "2",
    pages = "021001",
    year = "2023"
}

@article{Fiorillo:2023ytr,
    author = "Fiorillo, Damiano F. G. and Raffelt, Georg G. and Vitagliano, Edoardo",
    title = "{Large Neutrino Secret Interactions Have a Small Impact on Supernovae}",
    eprint = "2307.15115",
    archivePrefix = "arXiv",
    primaryClass = "hep-ph",
    doi = "10.1103/PhysRevLett.132.021002",
    journal = "Phys. Rev. Lett.",
    volume = "132",
    number = "2",
    pages = "021002",
    year = "2024"
}

@article{Fiorillo:2023cas,
    author = "Fiorillo, Damiano F. G. and Raffelt, Georg G. and Vitagliano, Edoardo",
    title = "{Supernova emission of secretly interacting neutrino fluid: Theoretical foundations}",
    eprint = "2307.15122",
    archivePrefix = "arXiv",
    primaryClass = "hep-ph",
    doi = "10.1103/PhysRevD.109.023017",
    journal = "Phys. Rev. D",
    volume = "109",
    number = "2",
    pages = "023017",
    year = "2024"
}

@article{Kazanas:2014mca,
    author = "Kazanas, Demos and Mohapatra, Rabindra N. and Nussinov, Shmuel and Teplitz, Vigdor L. and Zhang, Yongchao",
    title = "{Supernova Bounds on the Dark Photon Using its Electromagnetic Decay}",
    eprint = "1410.0221",
    archivePrefix = "arXiv",
    primaryClass = "hep-ph",
    reportNumber = "UMD-PP--014-015",
    doi = "10.1016/j.nuclphysb.2014.11.009",
    journal = "Nucl. Phys. B",
    volume = "890",
    pages = "17--29",
    year = "2014"
}

@article{Akita:2023iwq,
    author = "Akita, Kensuke and Im, Sang Hui and Masud, Mehedi and Yun, Seokhoon",
    title = "{Limits on heavy neutral leptons, Z' bosons and majorons from high-energy supernova neutrinos}",
    eprint = "2312.13627",
    archivePrefix = "arXiv",
    primaryClass = "hep-ph",
    reportNumber = "CTPU-PTC-23-55",
    doi = "10.1007/JHEP07(2024)057",
    journal = "JHEP",
    volume = "07",
    pages = "057",
    year = "2024"
}

@misc{JankaWeb,
  title = {Garching Core-Collapse Supernova Research Archive},
  howpublished = {\url{https://wwwmpa.mpa-garching.mpg.de/ccsnarchive/}}
}

\onecolumngrid
\appendix

\setcounter{equation}{0}
\setcounter{figure}{0}
\setcounter{table}{0}
\setcounter{page}{1}
\makeatletter
\renewcommand{\theequation}{S\arabic{equation}}
\renewcommand{\thefigure}{S\arabic{figure}}
\renewcommand{\thepage}{S\arabic{page}}

\begin{center}
\textbf{\large Supplemental Material for the Letter\\[0.5ex]
{\em Energy transfer by feebly interacting particles in supernovae: the trapping regime}}
\end{center}

\bigskip

In the Supplemental Material, we recollect standard results on the production of axions coupling to photons, provide an additional derivation of the approximate heat flux driven by axions, and show our constraints obtained integrating only for axions emitted after the end of the accretion phase.

\section{A.~Production of axions coupled to photons.}

The axion production can be expressed entirely in terms of the reduced absorption coefficient of an axion with energy $E_a$. The definition of this coefficient is most easily expressed by introducing the phase-space distribution for the axion $f_a$, defined so that $\int f_a d^3\mathbf{p}/(2\pi)^3=n_a$ is the axion number density and $\bf p$ is the axion momentum. The collisional evolution of the axion is described by a kinetic equation of the form
\begin{equation}
    \frac{\partial f_a}{\partial t}=Q_a(1+f_a)-\gamma_a f_a=Q_a-\tilde{\gamma}_a f_a,
\end{equation}
where $Q_a$ is the emission term, expressed in terms of the number of axions emitted per unit energy, time, and volume $dn_a/dE_a dt dV$ as 
\begin{equation}\label{eq:Q_equation}
    Q_a=\left(\frac{\partial f_a}{\partial t}\right)_{\rm em}=\frac{2\pi^2}{E_a \sqrt{E_a^2-m_a^2}}\frac{dn_a}{dE_a dt dV},
\end{equation}
while $\gamma_a$ is the absorption term. The reduced absorption coefficient is $\tilde{\gamma}_a$; from the law of detailed balance, we easily see that
\begin{equation}\label{eq:detailed_balance}
    \tilde{\gamma}_a=Q_a(e^{E_a/T}-1)=\frac{2\pi^2 (e^{E_a/T}-1)}{E_a\sqrt{E_a^2-m_a^2}}\frac{dn_a}{dE_a dt dV}.
\end{equation}
The axion luminosity, as defined in the main text, is therefore
\begin{equation}\label{eq:L_equation}
    \frac{dL}{dE_adR}=\frac{dn_a}{dE_a dt dV}E_a 4\pi R^2.
\end{equation}

It is therefore sufficient to specify the form of $\tilde{\gamma}_a$ to obtain directly the emissivity of the species. There are two contributing processes: inverse Primakoff $a N\to \gamma N$ and decay $a\to \gamma \gamma$. For inverse Primakoff, the absorption rate can be written as
\begin{equation}\label{eq:gammatilde_primakoff}
    \tilde{\gamma}_{a}=2\hat{n}\frac{k}{p}\frac{Z^2\alpha g_{a\gamma\gamma}^2}{2}f_P,
\end{equation}
where $k=\sqrt{E_a^2-\omega_P^2}$ is the momentum of the photon into which the axion is converted, $p=\sqrt{E_a^2-m_a^2}$ is the axion momentum, $Z$ is the atomic number (we take $Z=1$), $\hat{n}=Y_e n_B$ is the charge density (in our case $\hat{n}=n_e=n_p$ is equal to the electron and proton number density), and $g_{a\gamma\gamma}$ is the axion photon coupling. Here the plasma frequency is $\omega_P=4\alpha(\mu_e^2+\pi^2 T^2/3)/3\pi$, where $\mu_e$ is the electron chemical potential. Finally, we have
\begin{equation}
    f_P=\frac{[(k+p)^2+k_S^2][(k-p)^2+k_S^2]}{16k_S^2 kp}\log\frac{(k+p)^2+k_S^2}{(k-p)^2+k_S^2}-\frac{(k^2-p^2)^2}{16k_S^2 k p}\log\frac{(k+p)^2}{(k-p)^2}-\frac{1}{4}.
\end{equation}
Here, the Debye screening scale is $k_S^2=4\pi\alpha\hat{n}/T$, where $T$ is the temperature. In the absence of heavy elements, this gives
\begin{equation}
    \tilde{\gamma}_a=5.46\times 10^{18}\; \mathrm{s}^{-1}\;\frac{Y_e \rho}{10^{14}\;\mathrm{g/cm}^3}\;\left(\frac{g_{a\gamma\gamma}}{1\;\mathrm{GeV}^{-1}}\right)^2\frac{k}{p}f_P.
\end{equation}
Following Ref.~\cite{Caputo:2022mah}, we extract $Y_e$ from the simulation assuming that $Y_e=1-X_n$, where $X_n$ is the neutron fraction. Eq.~\eqref{eq:gammatilde_primakoff}, together with Eq.~\eqref{eq:detailed_balance}, reproduce Eq.~(S8) of Ref.~\cite{Caputo:2022mah}.

For the decay, we have
\begin{equation}
    \tilde{\gamma}_a=\frac{g_{a\gamma\gamma}^2 m_a^4 \tilde{f}_B}{64 \pi E_a},
\end{equation}
with
\begin{equation}
    \tilde{f}_B=\frac{2T}{p}\log\left[\frac{e^{\frac{E_a+p}{4T}}-e^{-\frac{E_a+p}{4T}}}{e^{\frac{E_a-p}{4T}}-e^{-\frac{E_a-p}{4T}}}\right].
\end{equation}

If $T\ll E_a, m_a$, this simplifies to $f_B\to 1$. We can also write
\begin{equation}
    \tilde{\gamma}_a=7.46\times 10^{18}\;\mathrm{s}^{-1}\;\left(\frac{g_{a\gamma\gamma}}{1\;\mathrm{GeV}^{-1}}\right)^2\;\left(\frac{m_a}{100\;\mathrm{MeV}}\right)^3 \frac{m_a}{E_a}f_B.
\end{equation}

\section{B.~Energy transport in the diffusive regime.}

In the main text, we have obtained an expression for the diffusive flux, in the form of Eq.~(11), which is valid in the regime $\lambda\ll R$, by expressing in terms of the thermal conductivity of the axions. On the other hand, its form looks significantly different from the expression in Eq.~(9), which is instead valid throughout all regimes. In this section, we show directly how Eq.~(9) may be approximated to reach the form of Eq.~(11). In the limit $\lambda\ll R$, only $L^{(1)}_{\rm PNS}$ and $L^{(1)}_{\rm prog}$ are non-vanishing in Eq.~(9);, since the very small mean free path impedes axions from leaving the progenitor star. In addition, because the mean free path is so small, all the factors related to spherical geometry can be neglected, since the emission of axion is completely local, at distances so small that the emitting surface can be approximated as an infinite plane. In other words, we can replace
\begin{equation}
    \tau(R, R_{\rm NS})\simeq \frac{z}{\lambda x}
\end{equation}
for $x>0$, since in Eq.~(2) in the main text $r\simeq R\simeq R_{\rm NS}$. Here we are denoting by $z=R_{\rm NS}-R$. Notice that for $x<0$ the optical depth is infinite; for a point close to the surface, the only way out of it, for a very small mean free path, is to move towards the surface. Similarly, Eq.~(7) and~(8) in this limit give $\tau_2\to +\infty$ and $\tau_1=-z/\lambda x$ for $x<0$. We may therefore write
\begin{equation}
    \frac{dL}{dE_a}=\int_0^{+\infty} dz \left.\frac{dL}{dE_a dR}\right|_{R=R_{\rm NS}-z}\int_0^{+1}\frac{dx}{2} e^{-\frac{z}{\lambda x}}-\int_{-\infty}^0 dz \left.\frac{dL}{dE_a dR}\right|_{R=R_{\rm NS}-z} \int_{-1}^0 \frac{dx}{2} e^{\frac{z}{\lambda x}},
\end{equation}
where the first term comes from $L^{(1)}_{\rm PNS}$ and the second one from $L^{(2)}_{\rm PNS}$.
By changing integration variable to $-z$ in the second term, we can rewrite it as
\begin{equation}
    \frac{dL}{dE_a}=\int_0^{+\infty}dz\left(\left.\frac{dL}{dE_a dR}\right|_{R=R_{\rm NS}-z}-\left.\frac{dL}{dE_a dR}\right|_{R=R_{\rm NS}+z}\right)\int_0^1\frac{dx}{2}e^{-\frac{z}{\lambda x}}.
\end{equation}
Since $\lambda$ is very small, the difference in parenthesis can be expanded for small $z$, allowing us to perform the integrals analytically
\begin{equation}
    \frac{dL}{dE_a}=-\frac{\lambda^2}{3}\frac{\partial}{\partial R}\left(\frac{dL}{dE_a dR}\right),
\end{equation}
where the derivative must be evaluated at the surface of the neutron star $R=R_{\rm NS}$. To proceed, we use Eqs.~\eqref{eq:Q_equation},~\eqref{eq:detailed_balance}, and~\eqref{eq:L_equation} to relate the luminosity to the mean free path
\begin{equation}
    \frac{dL}{dE_a dR}=4\pi R^2\frac{E_a p_a^2}{2\pi^2\lambda (e^{E_a/T}-1)}.
\end{equation}
In differentiating this with respect to $R$, we need to keep only the terms that contain the temperature gradient, since it is the latter that drives the heat flux, and it in this approximation that thermal conductivity emerges; the quality of this approximation is validated by the perfect matching with the numerical result of the complete expression in Eq.~(9). Thus, we finally find
\begin{equation}
    \frac{dL}{dE_a}=-\frac{2\lambda R_{\rm NS}^2 E_a^2 p_a^2 e^{E_a/T}}{3\pi T^2(e^{E_a/T}-1)^2}\frac{\partial T}{\partial R}.
\end{equation}
This reproduces completely the diffusive limit approximation obtained in Eq.~(11) by the heat transport argument.

\section{C.~Constraints from energy deposition after explosion.}

\begin{figure*}
    \includegraphics[width=\textwidth]{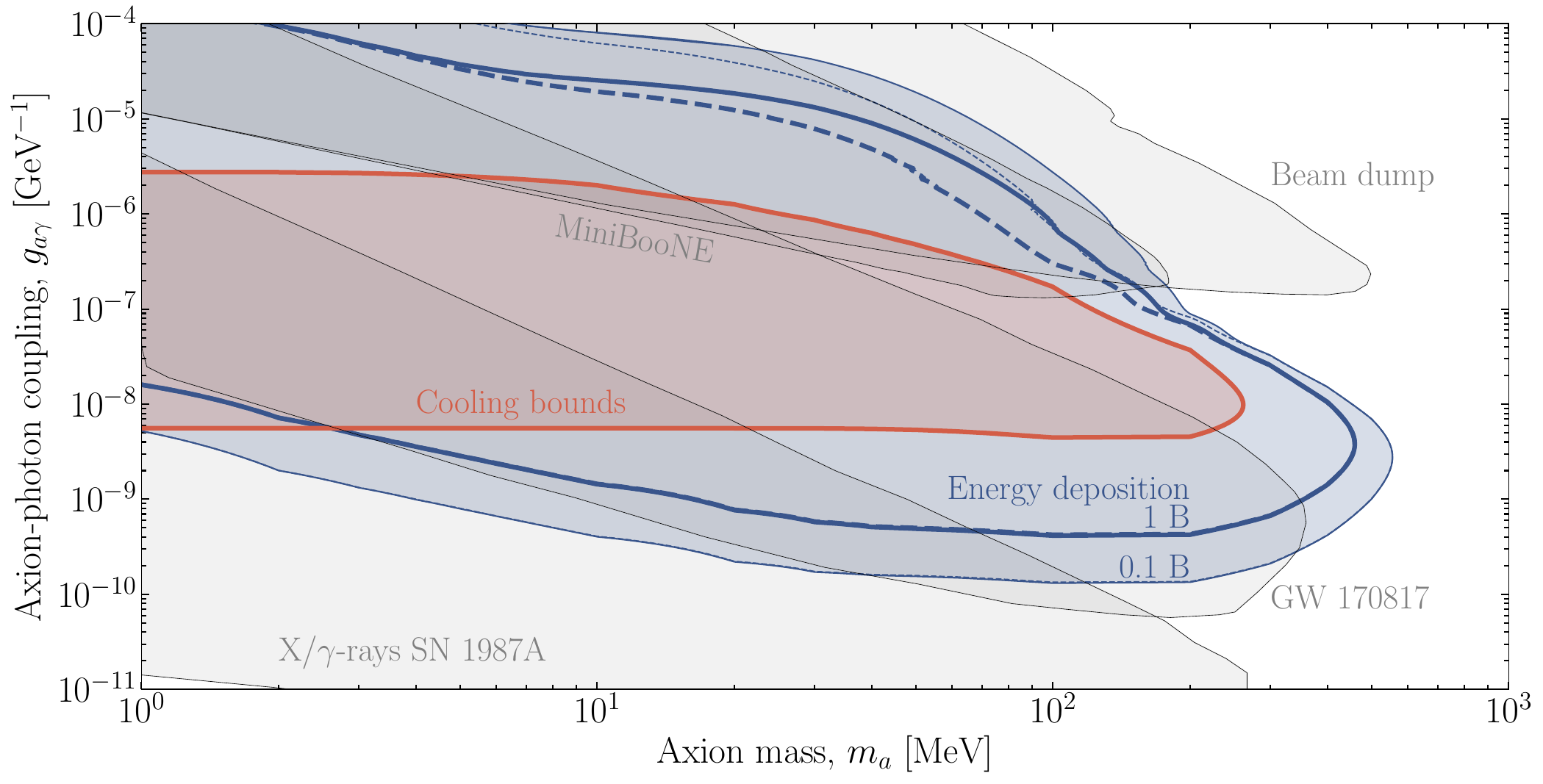}
    \caption{Constraints on axion-photon coupling accounting only for the energy deposited outside of the PNS after 300~ms (dashed), compared with the constraints shown in the main text (solid).}\label{fig:SM}
\end{figure*}

The energy extracted by axions from the PNS in the trapping regime remains as a heat flux just outside the PNS and is ultimately converted into photons at radii where the temperature is much lower than the axion mass. If deposition occurs within material with a sufficiently high infall velocity, some of this energy could dynamically return to the PNS, raising the concern that the constraints presented in the main text may be overly optimistic. In the free-streaming regime, this issue is generally negligible since axions decay at much larger distances from the PNS.

In reality, estimating the fraction of deposited energy that might be entrained by infalling material requires a fully self-consistent simulation of the deposition process. To adopt a very conservative approach, we assess how the constraints change when considering only energy deposition occurring after 300 ms post-bounce, approximately marking the end of the accretion phase (see, e.g., Table I in Ref.~\cite{Fiorillo:2023frv}).

Fig.~\ref{fig:SM} illustrates the impact of this choice. The free-streaming regime remains unaffected, as volumetric emission is largely dominated by later times, around 1s post-bounce. However, the axion heat conductivity at the PNS surface, which dictates the heat flux extracted from the PNS, is highly sensitive to the surface temperature, which drops rapidly. As a result, when the constraints become dominated by the diffusive regime, they can weaken by more than a factor of 2. We emphasize that, within the limitations of a post-processing approach, this is an overly conservative estimate, as a significant fraction of the energy deposited within the first 300~ms may still contribute to the explosion.



\end{document}